\documentclass[bibyear]{aa}  

\usepackage{graphicx}
\usepackage{txfonts}
\begin{document} 
\newcommand{\pks}{{\it PKS  0521-365}}
\newcommand{\mujy}{$\mu$Jy}
\newcommand{\convolution}{\mbox{$\ast$}}
\newcommand{\degree}{\mbox{$^\circ$}}

   \title{Anatomy of a blazar  in the (sub-)millimeter: ALMA Observations of \pks}

   \author{S. Leon\inst{1},
          P. C. Cortes\inst{1,5},
          M. Guerard\inst{1},
          E. Villard\inst{1},
          T. Hidayat\inst{2}, 
          B. Oca\~na Flaquer\inst{3},
          B. Vila-Vilaro\inst{1}           
          }

   \institute{$^1$ Joint ALMA Observatory - ESO, Av. Alonso de C\'ordova, 3104, Santiago, Chile \\
   $^2$ Bosscha Observatory and Astronomy Research Division, FMIPA, Institut Teknologi Bandung, Indonesia \\
   $^3$ CAUP, Porto University, Portugal \\
   $^4$ Instituto de Astrof\'{i}sica de Andaluc\'{i}a-CSIC, Granada, Spain \\
   $^5$ National Radio Astronomy Observatory, 520 Edgemont Rd, Charlottesville, VA, 22903, USA 
   }

   \date{Received September 15, 1996; accepted March 16, 1997}

 
  \abstract
   {}
   {We aim at analyzing the (sub-)millimeter emission in a nearby blazar, \pks, 
to study the synchrotron and thermal emission in
the different components detected at low frequency.}
{We analyze the archive public  data of the  ALMA Cycle 0  where \pks~ is used as a calibrator. 
A total of 13 projects with 23  dataset is analyzed  in band 3, 6 and 7 and combined.
The whole set of data is combined and wavelet filtered to obtain a deep image towards \pks~ reaching
   a dynamic range of 47000. The individual emission flux is  measured at different date over a period of
 11 months in various components. Finally we analyze the Spectral Energy Distribution (SED) in  each different component, including the radio jet and counter jet.
}
{The point sources detected in the field follow a similar distribution to  previous studies. The blazar flux shows large variation especially in band 3. 
Different components are observed: core, 
radio jet and newly detected counter jet, Hot Spot (HS) and a disky structure roughly perpendicular to the jet. 
The HS emission is formed by a point source  surrounded by an extended emission.
The viewing angle of the jet is about $30^\circ$ with a Doppler factor of $\delta = 1.6$. 
The HS is at a distance of 19 kpc from the center. The SED analysis shows a strong variation of 
the core spectral index, especially in band 3. The two components in the radio jet have  roughly 
a flat spectral index in band 6 and 7.}
{Using these ALMA data the different weak and extended components in 
\pks~ are detected. The analysis of both jets constrains the geometrical distance of the HS to the center. 
The SED presents a different shape in time and frequency for each component. 
Finally a new structure is detected roughly perpendicular to the radio jet and a thermal emission origin
is currently favoured. Further observations at higher spatial resolution  are needed to confirm 
that hypothesis.}


   \titlerunning{ALMA (sub-)millimeter study of \pks}
   \authorrunning{Leon et al.}
   \maketitle
%

\section{Introduction}

Inside the large range of different type of active galactic nuclei (AGN) 
blazars are the most powerful sources. Showing intense emission across the
electromagnetic spectrum, it is thought that
this strong continuum emission is due to an ejection of relativistic plasma nearly on the line of sight. 
A radio jet is commonly observed in blazars at lower frequencies 
(Antonucci \& Ulvestad, 1985) and their strong  flux is accounted for the relativistic beaming. 
A blazar is then a particular case of a
head-on radio galaxy from an underlying population with a spread geometry distribution of the 
jet (Antonucci \& Ulvestad, 1985; Urry \& Padovani 1995) originating
from an AGN and oriented nearly head-on towards the observer according to the
 unified scheme proposed two decades ago (e.g. Urry \& Padovani, 1995). 
The jet can be detected up to large energy ranges (Gamma, X-ray) because of the strong beaming effect. 
Blazars can be divided into two types, the low power AGNs, mainly without Broad-Line Regions (BLRs) and a large 
synchrotron break frequency (Capetti et al. 2010, Tavecchio et al. 1998), and the
Flat-Spectrum Radio Quasars (FSRQs) with a lower break frequency (Maraschi \& Tavecchio 2003). 
The first kind forms the family of BL Lac radio galaxies which was the first
 of its kind detected (e.g.  Blandford \& Rees 1978). 
The SED of the blazar shows normally two bumps, 
the synchrotron bump at low energy peaking in the
infrared (IR)-X-ray and the Inverse Compton (IC) bump normally in $\gamma$-ray. 
Each bump is characterized by a peak frequency $\nu_1$ and $\nu_2$ at the peak
of the emission in each regime. Fossati et al. (1998) found an anti-correlation 
between the peak frequency $\nu_1$ and the bolometric and synchrotron luminosity.
They suggested a link between the source power and the frequency peak. 
The sequence may match a spectroscopic one, from powerful FSRQ sources at the low-peak,
high-luminosity end and the BL Lacs with high-peak and low-luminosity. 
The BL Lacs do not show normally signs of BLRs. It was suggested that the Inverse Compton (IC) emission in
FSRQ was produced through the Synchrotron Self Compton (SSC) mechanism or from external 
seed photons from the BLRs. These photons would be up-scattered by the
synchrotron electrons (e.g. Chen et al. 2012). The long-standing Unified Scheme for radio-loud AGN proposes that the 
Fanaroff-Riley I and II radio galaxies (Fanaroff \& Riley, 1974) 
are the underlying progenitors of BL Lacs and FSRQs galaxies (Bicknell 1995, Fossati et al. 1998).

In a population synthesis study, Lister \& Marscher (1997) used realistic
luminosity functions to show that the majority of compact extragalactic radio sources
have jets with low Lorentz factors.
The AGN in general and the jet in particular are known to drive the evolution of their host galaxy (Best \&  Heckman 2012, 
Khalatyan et al. 2008, Sturm et al. 2011) through a feedback mechanism of depositing energy
in the interstellar medium (ISM). At large scale the radio jet can slow down or stop the cooling of the 
massive X-ray halo (Fabian 2012) that surrounds these galaxies. At the galactic scale 
the jet interacts with the local ISM (molecular gas, X-ray halo, etc), and it has been 
suggested that for host galaxies with a dense ISM phase the radio jet could
be completely confined to the central part with no outwards propagation (Carvalho 1998) 
forming Compact Steep Spectrum galaxies. 
It was argued recently (Rashed
et al., 2013) that the backflow along the radio jet in the elliptical host galaxy 
SDSS J080800.99+483807.7 might have triggered star formation. This star formation triggering mechanism
in radio galaxies has been used to explain 
the correlation between the radio and optical morphologies of distant 3C 
radio galaxies (McCarthy et al., 1987). 
Some of the most dramatic jet-induced star formation, with star formation rates (SFR) 
as high as 1000 $M_\odot$yr$^{-1}$, 
is associated with very luminous radio galaxies at redshifts up to $z\sim 4$ 
(Dey et al. 1997; Bicknell et al. 2000).

One study has found a clear interaction of the radio jet and the ISM in the Narrow Line Regions (NLRs) of Seyfert galaxies
(Falcke et al., 1998). Such interactions play a major role in determining the morphology of the NLR, as the radio jets sweep up and compress ambient gas, producing 
ordered structures with enhanced surface brightness in line emission. The nearby radio galaxy M 87 which can be classified as a BL Lac was recently found to have 
a giant  radio-to-X-ray synchrotron  (Stawarz et al., 2006). It was argued that the presence of a gaseous concentration in the central part would
provoke the interaction of the jet with such a feature resulting in the formation of a stationary shock structure in the innermost parts of the 
M 87 jet.

The radio source \pks~ (J0522-364) is a nearby AGN (Danziger et al. 1979) and
it is one of the most remarkable extragalactic objects of the southern sky,
because it exhibits a variety of nuclear and extra-nuclear phenomena.
In this work we assume a cosmology with  $\Omega_m = 0.27$, $\Omega_\Lambda = 0.73$, $H_0 = 71$ km s$^{-1}$ Mpc$^{-1}$ 
(Spergel et al. 2003). At $z = 0.0565$, 
\pks~  is at a comoving radial distance of 236 Mpc, and have a linear scale size of 1.15 kpc/arcsec.
Classified as a BL Lac object, it shows strong narrow and broad emission lines
typical of type
1 AGN both in the optical and the UV (Ulrich 1981; Danziger et al. 1983; Scarpa et al. 1995).
The source has a prominent radio, optical, and X-ray jet
(Danziger et al. 1979; Keel 1986; Falomo 1994;
Macchetto et al. 1991; Scarpa et al. 1999; Birkinshaw et al. 2002),
which resembles that of the nearby radio galaxy M 87 (Sparks et al. 1994).

\noindent Here we present archival ALMA observations of the radio source \pks.
We retrieved public Cycle-0 observation where \pks~ was used as calibrator in band 3, 6, and 7.
Section \ref{obs} presents the observations and data reduction, section \ref{ana} the 
millimeter and sub-millimeter distribution of the emission, section \ref{var} the variability
in the emission, section \ref{spindex} computes the spectral index, 
section \ref{discussion} the discussion, and section \ref{conclusions} the 
summary and conclusions.



\section{Observations and data reduction}
\label{obs}
Publicly available data from ALMA Cycle 0 observations 
were used to analyze the sub-millimeter properties of the blazar \pks. 
As shown on Table \ref{tab:obs-journal} 
the observations span a period of about 1 year from 2012/01/10 to 2012/11/19. 
The observed frequencies go from 87.3 GHz up to 372.5 GHz, which covers 
 bands 3, 6 and 7.
\pks~ was generally used as a bandpass calibrator for the science projects sharing the same spectral
setup, and observed 
between 2 to 4 minutes per execution. If two ExecBlocks (hereafter EBs) were observed
consecutively, we merged them in a single dataset and the observing time was chosen as  
the starting time of the first EB. Each EB has four
spectral windows of, generally, 2 GHz bandwidth distributed over 4 basebands, 
two in the USB window and two in the LSB window. 
For the flux determination of each dataset the 
four spectral windows were combined and the LSB/USB spectral windows were used to determine 
the spectral index $\alpha$, hereafter defined as $S_\nu \propto \nu^\alpha$ with $S_\nu$ the flux density 
at the frequency $\nu$, in one dataset as indicated further. 
Before combining the different dataset in each band, a self-calibration was 
performed on each of them to improve the phase/amplitude gain solutions thanks 
to the strong continuum emission. The effective frequency quoted for the combined bands was computed using the deconvolution CLEAN task in CASA with the multi-frequency 
synthesis (mfs) mode, see Rau  \& Cornwell (2011) for details.  

The two most prominent feature of \pks~ are the strong core and the so-called "Hot Spot" (HS) 
located at 8.6\arcsec at the South-East of the core. The main properties
of each dataset and the measured flux of the core and the Hot Spot are summarized in 
 Table \ref{tab:obs-properties}. For the core the flux in the central area was measured directly since the jet contribution 
 is less than 1\% of the total flux. The HS flux was measured including the pixels above 3-$\sigma$ of the RMS in the HS area.

\begin{table*}[htbp]
\caption{Journal of the observations}
\centering
\begin{tabular}{|l|l|l|l|l|}
\hline
Id  & Project UID & Date   & Frequency & Integration \\
    &             &        & (GHz)     &  (seconds)       \\
\hline
J0522-364\_99\_1      &  2011.0.00099.S      & 2012/04/06 22:19  & 98.7/110.3  &   123 \\
J0522-364\_99\_2      &  2011.0.00099.S       & 2012/07/04 11:53  & 98.7/110.3  &   110 \\
J0522-364\_99\_3      &  2011.0.00099.S       & 2012/07/03 11:25  & 100.6/112.5 &   171     \\ 
J0522-364\_131\_1     &  2011.0.00131.S       & 2012/01/24 23:20  & 101.4/114.5 &   296 \\
J0522-364\_467\_1     &  2011.0.00467.S      & 2012/03/27 21:23  & 96.5/108.3  &   175 \\ 
J0522-364\_108\_3     &  2011.0.00108.S       & 2012/07/29 10:35  & 87.3/99.0   &   233   \\
\hline
J0522-364\_268\_1     &  2011.0.00268.S       & 2012/01/10 21:55   & 252.8/241.3 &   200 \\
J0522-364\_268\_2     &  2011.0.00268.S      & 2012/01/13 02:26   & 252.8/241.3 &   200  \\
J0522-364\_210\_1\_2  &  2011.0.00210.S       & 2012/08/09 09:58   & 219.7/231.2 &   463   \\
J0522-364\_210\_3     &  2011.0.00210.S       & 2012/08/26 09:14  & 219.7/231.2 &   236   \\ 
\hline
J0522-364\_108\_1     &  2011.0.00108.S       & 2011/11/05 06:51   & 342.8/354.3  &  115  \\
J0522-364\_108\_2     &  2011.0.00108.S       & 2011/11/06 02:05   & 342.8/354.3  &   115   \\
J0522-364\_397\_1     &  2011.0.00397.S      & 2011/11/16 05:24   & 339.0/351.0 &  141 \\
J0522-364\_124\_1\_2  &  2011.0.00124.S      & 2012/07/18 09:41   & 318.1/330.1 &  464    \\ 
J0522-364\_768\_1     &  2011.0.00768.S      & 2012/07/31 09:54   & 339.8/351.7 &   175   \\ %
J0522-364\_768\_2\_3  &  2011.0.00768.S      & 2012/08/14 09:15   & 339.8/351.7 &  351    \\ 
J0522-364\_83\_1      &  2011.0.00083.S     & 2012/06/16 11:25  & 342.8/354.7 &  326    \\
J0522-364\_83\_2      &  2011.0.00083.S     & 2012/06/17 11:43  & 342.8/354.7 &  326    \\
J0522-364\_83\_3      &  2011.0.00083.S      & 2012/08/27 06:59  & 342.8/354.7 &  326    \\ 
J0522-364\_208\_1     &  2011.0.00208.S     & 2012/10/22 03:56  & 342.4/354.4 &  200      \\
J0522-364\_223\_1     &  2011.0.00223.S      & 2012/10/21 06:21  & 330.8/345.6 &  342        \\
J0522-364\_223\_2     &  2011.0.00223.S      & 2012/10/22 05:08  & 330.8/345.6 &  342        \\
J0522-364\_340\_1\_2  &  2011.0.00340.S      & 2012/11/19 07:36  & 358.0/372.5 &  130          \\
\hline
\end{tabular}

\label{tab:obs-journal}
\end{table*}

\begin{table*}[htbp]
\caption{Dataset properties (natural weighting)}
\centering
\begin{tabular}{|l|l|l|l|l|l|l|}
\hline
Id  &  MJD &  $\nu_{eff}$  &    Beam  &  RMS  & Flux (Core)  & Flux (HS) \\
    &      &      (GHz)      &   (arcsec)     &   (mJy)    &  (mJy)  &  (mJy)       \\
\hline
J0522-364\_99\_1  & 56023.93   & 104.55 &    $1.34\times 1.72$     & 3.9  &  7174       &     83       \\
J0522-364\_99\_2  & 56112.50  & 104.33 &    $1.69\times 2.41$     & 3.0  &  6111       &    178        \\
J0522-364\_99\_3  & 56111.48   & 106.62 &    $1.43\times 2.46$     & 0.6  &  4858       &    163        \\
J0522-364\_131\_1 & 55950.97   & 107.99 &    $3.14\times 4.46$     & 1.4  &  5317       &    142        \\
J0522-364\_467\_1 & 56013.89   & 102.40 &    $1.47\times 1.69$     & 3.0  &  9623       &    133        \\
J0522-364\_108\_3 & 56137.44   & 93.16  &    $1.52\times 2.64$     & 1.3  &  5474       &    177        \\
\hline
J0522-364\_268\_1 & 55936.91   & 247.08 &    $1.32\times 2.68$    &  2.1  &  4208      &  33          \\
J0522-364\_268\_2 & 55939.10   & 247.08 &    $1.19\times 1.78$    &  1.0  &  4524      &  44          \\
J0522-364\_210\_1\_2 & 56148.42 & 225.43 &    $0.67\times 0.91$   &  0.3  &  4732      &  53    \\
J0522-364\_210\_3 & 56165.38   & 225.44 &    $0.68\times 0.97$    &  0.7  &  4735      &  52      \\ 
\hline
J0522-364\_108\_1  & 55870.29  & 348.58 &    $1.18\times 1.35$    &  0.7  & 5345        &  17        \\
J0522-364\_108\_2  & 55871.09  & 348.58 &    $1.17\times 2.95$    &  0.6  & 5077        &  14        \\ %
J0522-364\_397\_1  & 55881.22  & 344.99 &    $1.23\times 1.39$    &  0.9  & 5254        &    19      \\
J0522-364\_124\_1\_2 & 56126.40 & 324.15 &    $0.43\times 0.80$    &  0.3  & 3310        &    25       \\
J0522-364\_768\_1  & 56139.41  & 345.77 &    $0.41\times 0.75$    &  1.8  & 3973        &    22       \\   
J0522-364\_768\_2\_3 & 56153.39 & 345.77 &    $0.45\times 0.66$    &  0.2  & 4087        &    23      \\   
J0522-364\_83\_1   & 56094.48  & 348.82 &    $0.58\times 0.91$    &  0.6  & 4870        &    21       \\  
J0522-364\_83\_2   & 56095.49  & 348.82 &    $0.47\times 0.82$    &  0.8  & 4683        &    26       \\   
J0522-364\_83\_3   & 56166.29  & 348.82 &    $0.44\times 0.95$    &  4.1  & 3733        &    31        \\  
J0522-364\_208\_1  & 56222.16  & 348.42 &    $0.52\times 0.79$    &  0.5  & 4245        &    26        \\  
J0522-364\_223\_1  & 56221.26  & 338.19 &    $0.51\times 0.60$    &  0.3  & 4518        &    24        \\  
J0522-364\_223\_2  & 56222.21  & 338.19 &    $0.52\times 0.63$    &  0.4  & 4588        &    19         \\ 
J0522-364\_340\_1\_2 & 56250.32  & 365.28 &    $0.50\times 0.61$    &  0.9  & 7666        &    10         \\ 

\hline
\end{tabular}

\label{tab:obs-properties}
\end{table*}

The self-calibrated datasets were combined in each band and imaged. 
A final superset  of all self-calibrated dataset, called "band 3+6+7", was created and imaged to produce
the most sensitive continuum image at an effective  frequency of 221.02 GHz. 
The total integration time is of 1.55 hours with an equivalent of 66  pad positions
and 2145 baselines. The self-calibration of these data allowed us to achieve a RMS noise of 
158 \mujy\ for the band 3+6+7 with a dynamic range (SNR) of 27100 without wavelet filtering. The final spatial
resolution with a natural weighting is of $0.53\arcsec\times 0.71\arcsec$. It is worth noting that the band 7 combined image has the deepest sensitivity (91 \mujy) which is similar to 
the all-in-one combined data. The properties of the combined dataset in band 3, 6, 7 and 3+6+7 are summarized in Table \ref{tab:obs-combine} together with the 
Hot Spot flux which is not varying with time. Each band reaches an effective bandwidth of 27 GHz (band 3) to 285 GHz (band 3+6+7). The effects
of the lowest spatial frequencies observed at the different bands was checked by computing the flux of the HS in band 3 using the 
spatial frequencies of band 7. The HS flux is in that case lower by 5\% compared to the  measured flux with the full uv-coverage. 
Given the calibration
uncertainty we can assume safely that the differential sensitivity to spatial scales between bands is negligible in the case of the observations of
\pks .

\begin{table*}[htbp]
\caption{Sets of combined data (natural weighting) together with the Hot Spot flux. The properties indicated are with the wavelet filtering. 
See text for more details.}
\centering
\begin{tabular}{|c|c|c|c|c|c|}
\hline
Set  &  $\nu_{eff}$  &    Beam  &  RMS  & Dynamic Range  & Flux (HS)  \\
    &    (GHz)      &   (arcsec) &  ($\mu$Jy) &       &  (mJy)       \\
\hline	
band 3  &   100.43  &  $1.59\times 2.49$ (PA=78\degree) &    306     & 15300    &   153     \\     
band 6  &   237.06  &  $0.69\times 0.93$ (PA=93\degree) &    147     & 28500   &   39      \\    
band 7  &   336.44  &  $0.48\times 0.64$ (PA=90\degree)  &    91     & 40000   &   22      \\     
band 3+6+7 & 221.02 &  $0.53\times 0.71$ (PA=90\degree)  &    90     & 47000   &   43      \\     
\hline
\end{tabular}

\label{tab:obs-combine}
\end{table*}

\subsection{Wavelet filtering}

The wavelet transform is a powerful signal processing technique which provides a decomposition of the signal into elementary local contribution labeled 
by a scale parameter (Grossman \& Morlet 1985). They are the scalar products with a family of shifted and dilated functions of constant shape called
wavelets. The data are unfolded in a space-scale representation which is invariant with respect to dilation  of the signal. Such an analysis is 
particularly suited to study signals which exhibit space-scale discontinuities  and/or hierarchical features. \\
The wavelet analysis of the \pks~ datasets is performed on the combined dataset in each band after the deconvolution process using the CLEAN algorithm with natural
weighting.  As described in Leon et al. (2000) we use the "A trous" algorithm (see  Bijaoui 1991).   It  allows to get  a
discrete wavelet decomposition   within  a reasonable CPU time.    The kernel function  $B_s(x,y)$  for the  convolution   is a $B_3$  spline
function.  

\noindent The  wavelet transform (WT) $W(i,x,y)$  of the image $I(x,y)$  is then  obtained from the 
following steps:
\begin{eqnarray}
c_o(x,y) & = & I(x,y), \\
c_i(x,y)& = & c_{i-1} \convolution B_s(\frac{x}{2^i},\frac{y}{2^i}), \\
W(i,x,y)& = & c_i(x,y)-c_{i-1}(x,y).  
\end{eqnarray}
The last plane, called Last Smoothed Plane  (LSP), is the residuals of the last convolution  and not a wavelet  plane, but for convenience we will
  speak of wavelet  plane for  all  these planes. Each  plane $W(i,x,y)$ represents  the details of the  image at the   scale $i$. A wavelet plane
  $i$ has a  spatial resolution of $0.86\times 2^i$ pixels which translates into 0.21\arcsec\ in band 3 and 0.08\arcsec\ in band 6, 7 and combined 
  for the wavelet plane 0.

\noindent Then each   component of the  raw  map $W(i,x,y)$ was filtered, above a  given threshold   $\beta$, using the RMS  noise  $\sigma_i$  on  each
wavelet scale  $i$ to get the filtered  wavelet planes $W_f(i,x,y)$ of the radio continuum distribution in the different bands:
\begin{eqnarray*}
W_f(i,x,y) & = & W(i,x,y)  \mbox{\small \hspace{0.1cm}  
                 If $|W(i,x,y)| > \beta \sigma_i$} \\
           & = & 0         \mbox{\small  \hspace{1.4cm} Otherwise}       
\end{eqnarray*}

\noindent In this study the wavelet coefficients are filtered  at the 4.5-$\sigma$ level. After WT filtering for the continuum distribution  in the different bands
the actual dynamic range increases in the different filtered  set of data. Thus the combined band 3+6+7, after WT filtering, 
has a dynamic range of 47000 and  the combined band 7 reaches a dynamic range of 40000. By combining the wavelet scales one can achieve a different spatial resolution
with a specific RMS noise. Since the contribution of an image $I$ is spread over the different spatial scales, the reconstruction through a limited set of wavelet
scales is a lower limit to the total emission contribution. That effect is visible in  Table \ref{tab:wavelet-filtering} where the core flux is computed for the 
different wavelet reconstructions. A final equivalent RMS noise of 90 \mujy\ can be achieved for the band 3+6+7. It is worth noting that the "A trous" wavelet decomposition
is not strictly orthogonal and the spatial scale contribution can be spread over different wavelet scales. But the algorithm has a quasi-isotropic kernel which makes 
the physical analysis of each scale easier to perform. In our notation of the WT for the continuum images the scale 0 has the highest spatial resolution and 
the scale 7 is the LSP. The maximum scale was chosen to be 7 since largest wavelet scales are not giving fundamental informations.

\begin{table*}[htbp]
\caption{Properties of different wavelet reconstruction of the band 3+6+7 combined and WT filtered at 4.5-$\sigma$. In our notation the scale 7 is 
the LSP (see text).}
\centering
\begin{tabular}{|l|l|l|}
\hline
Wavelet Planes   & RMS       &  Core Flux  \\
                 & ($\mu$Jy) &     (\%)       \\
\hline
6 - 7            &  3.7       &  60    \\
5 - 7            &  17.3      &  68     \\
4 - 7            &  38.9      &  77     \\
3 - 7            &  64.8      &  86     \\
2 - 7            &  99.3      &  95     \\
1 - 7            & 107.9      &  100   \\
\hline
\end{tabular}

\label{tab:wavelet-filtering}
\end{table*}

\section{(Sub-)millimeter distribution}
\label{ana}
\begin{figure}
\includegraphics[width=9cm,height=8cm]{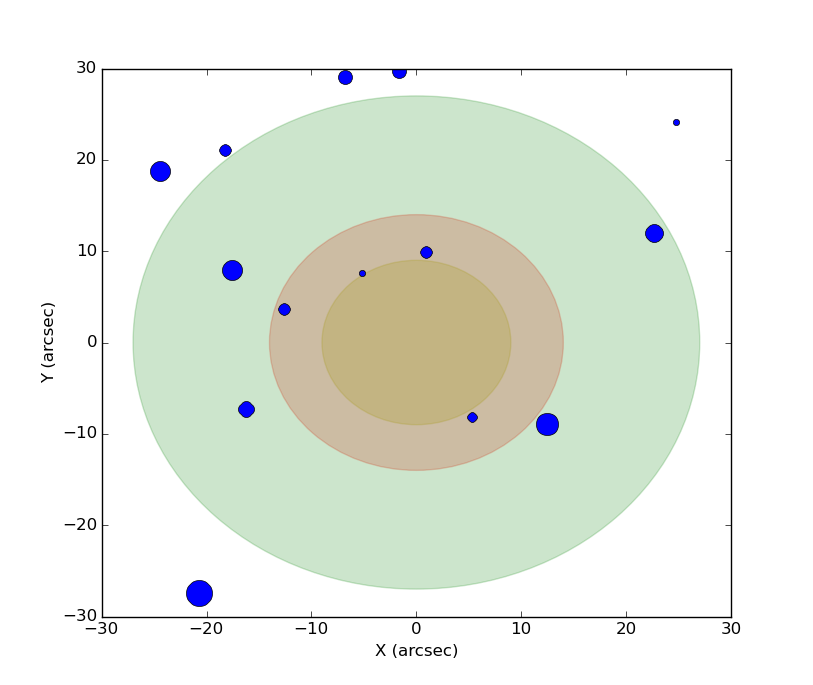}
\caption{Point sources detected in the combined field of band 3, 6 and 7 ($\nu_{eff} = 221.02$ GHz) after WT filtering at 4.5$\sigma$. 
The symbols are proportional to the flux. The field is centered on \pks. The colored circles represent the primary beamsize of each band.}
\label{fig:B367-PS}
\end{figure}

\begin{figure}
\includegraphics[width=9cm,height=7cm]{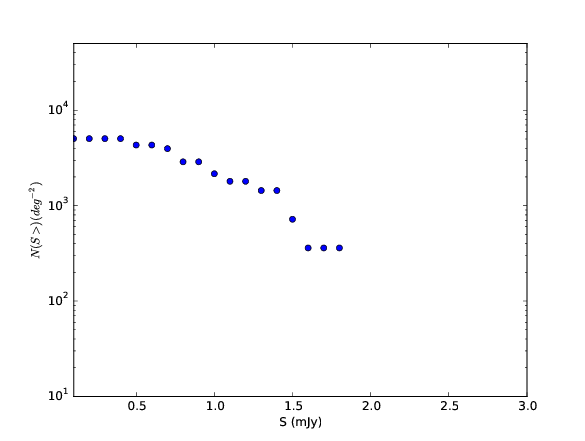}
\caption{Cumulative number count at 221 GHz in a field of $1\arcmin\times 1\arcmin$ between a flux of 0.1 mJy and 3 mJy. The effective RMS noise in
the WT-filtered image is  100 \mujy.}
\label{fig:source-cumulative-count}
\end{figure}

\begin{figure}
\includegraphics[width=9cm,height=9cm]{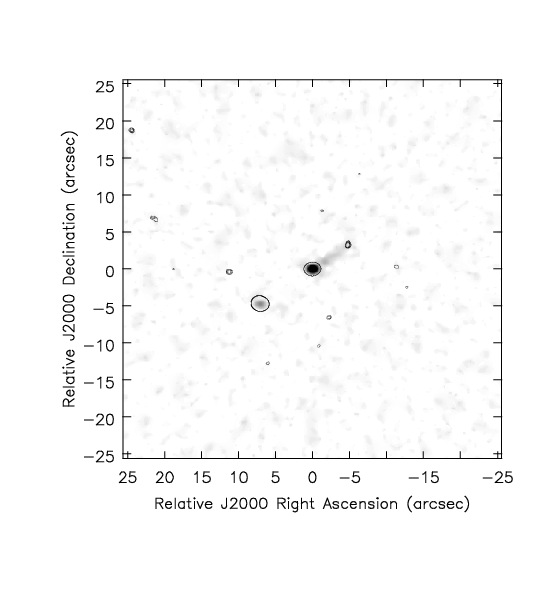}
\caption{Contour levels at [4, 4.5, 5, 5.5, 6]-$\sigma$ of the B6+B7 WT for the scales 2 to 4. The image is the band 7 emission.}
\label{fig:B6B7-2-4}
\end{figure}

In order to analyze the sub-millimeter emission towards \pks~ the point source distribution was searched in that field. The point sources were searched in the 
band 3+6+7 WT-filtered field with an effective RMS noise of 90 \mujy. The list of the point sources was obtained by using the SExtractor software (Bertin \& Arnouts, 
1996) at a 4.5-$\sigma$ level 
with a minimum of 5 pixels per sources to avoid spurious detections.
A total of 16 sources are detected in a $1\arcmin\times1\arcmin$ box and their properties are listed in Table \ref{tab:ps-properties}. Few sources are associated with the
continuum emission in \pks~ and most of them should be related with background sources as shown on Figure
 \ref{fig:B367-PS}. A comparison of the maximum peaks  detected 
in the psf image was done with the point sources and no correlation was found. The source \#10 and \#12 are respectively associated with the core and the HS emission.
It is important to note that the spatial distribution of these sources are not clustered on the radio continuum emission of \pks~ but rather uniform through the band 3+6+7
field. Because of the RMS noise of 90  \mujy\ 
in the WT-filtered band 3+6+7 field we can estimate the number of sources with a flux greater than 0.5 mJy to be $S(>0.5 \mbox{mJy}) \sim 5760 \deg^{-2}$. An estimation
of the cumulative count at 221 GHz is given on Figure \ref{fig:source-cumulative-count} between a flux of 0.1 mJy and 3 mJy. This estimation is not addressing entirely
the completeness of the detection and the possible spurious detections and bias. But we note that this estimation is similar, but lower,  to the one derived by Hatsukade 
et al. (2013) using ALMA fields at 238 GHz. Figure \ref{fig:B6B7-2-4} shows the band 6 and 7 emission by selecting the wavelet planes 2 to 4 which biases the slightly 
extended emission. The analysis of the point sources is beyond the scope of this article.

\begin{table*}[htbp]
\caption{Properties of the point sources detected in the band 3+6+7 ($\nu_{eff} = 221.02$ GHz) at a 4-$\sigma$ level in a box of $1\arcmin\times1\arcmin$. The $X,Y$ 
spatial positions are relative to the nucleus of \pks. The spatial  resolution of the band 3+6+7 dataset is $0.53\arcsec\times 0.71\arcsec$. }
\centering
\begin{tabular}{|l|l|l|l|l|l|l|l|l|}

\hline
Id   & $\alpha_{2000}$ & $\delta_{2000}$ & X  & Y  & Flux & Flux Error & FWHM    \\
     &  (degree)      &  (degree)       & (\arcsec) & (\arcsec)  & (mJy)  & (mJy) & (\arcsec)   \\
\hline
1  & 80.7422245  & -36.4503386 &  -1.7 &  29.7 &   1.05  &  0.05  &  0.59 \\
2  & 80.7440029  & -36.4505140 &  -6.9 &  29.1 &   1.06  &  0.05  &  0.51 \\
3  & 80.7331005  & -36.4518921 &  24.7 &  24.1 &   0.58  &  0.03  &  0.35 \\
4  & 80.7479270  & -36.4527443 & -18.2 &  21.1 &   0.80  &  0.04  &  0.44 \\
5  & 80.7500829  & -36.4533700 & -24.5 &  18.8 &   1.50  &  0.05  &  0.55 \\
6  & 80.7338043  & -36.4552554 &  22.7 &  12.0 &   1.39  &  0.06  &  0.66 \\
7  & 80.7413304  & -36.4558385 &   0.9 &   9.9 &   0.85  &  0.04  &  0.49 \\
8  & 80.7476982  & -36.4563809 & -17.5 &   8.0 &   1.50  &  0.06  &  0.62 \\
9  & 80.7434403  & -36.4564975 &  -5.2 &   7.6 &   0.57  &  0.03  &  0.35 \\
10  & 80.7415967  & -36.4585668 &   0.1 &   0.1 &  4302.94  &  0.07  &  0.66 \\
11  & 80.7460065  & -36.4575836 & -12.7 &   3.6 &   0.84  &  0.04  &  0.44 \\
12  & 80.7440576  & -36.4598546 &  -7.0 &  -4.5 &  45.47  &  0.11  &  0.98 \\
13  & 80.7472396  & -36.4606248 & -16.2 &  -7.3 &   1.12  &  0.05  &  0.51 \\
14  & 80.7397929  & -36.4608532 &   5.3 &  -8.1 &   0.72  &  0.04  &  0.35 \\
15  & 80.7373352  & -36.4610799 &  12.5 &  -8.9 &   1.64  &  0.05  &  0.55 \\
16  & 80.7487821  & -36.4662021 & -20.7 & -27.4 &   1.97  &  0.07  &  0.87 \\
\hline 
\end{tabular}

\label{tab:ps-properties}
\end{table*}

\begin{figure*}
\includegraphics[width=12cm]{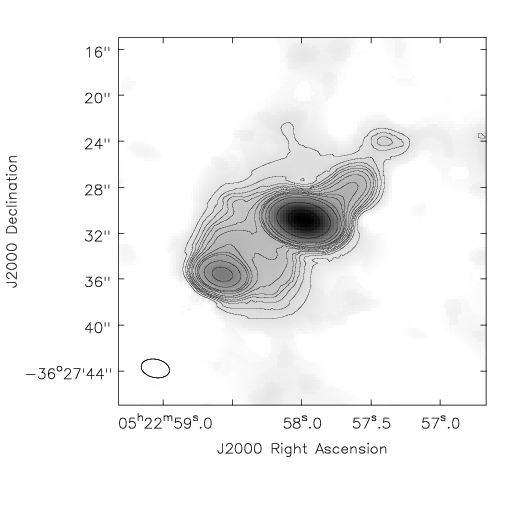}
\caption{Continuum emission towards \pks\ in band 3 ($\nu_{eff} =100.43$ GHz) after filtering at 4.5-$\sigma$ using a WT. The contours are 
at [3, 5, 7, 9, 11, 15, 20, 30, 40, 50, 100, 200, 1000, 5000]-$\sigma$ for a RMS noise of 306 \mujy.}
\label{fig:B3-wt}
\end{figure*}

\begin{figure*}
\includegraphics[width=12cm]{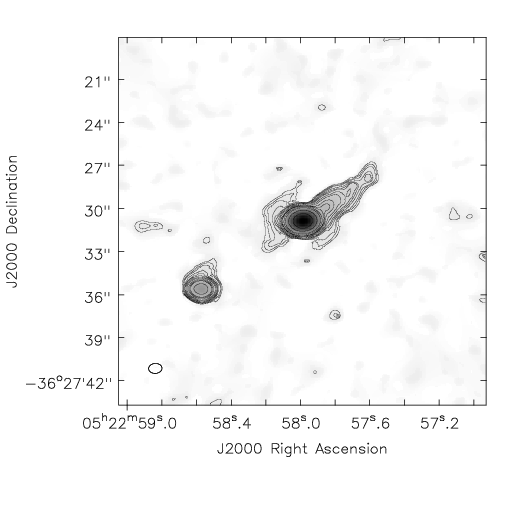}
\caption{Continuum emission towards \pks\ in band 6 ($\nu_{eff} = 237.06$ GHz) after filtering at 4.5-$\sigma$ using a WT. The contours are 
at [3, 4, 5, 7, 9, 11, 15, 20, 30, 40, 50, 100, 200, 1000, 5000]-$\sigma$ for a RMS noise of 147 \mujy.}
\label{fig:B6-wt}
\end{figure*}

\begin{figure*}
\includegraphics[width=12cm]{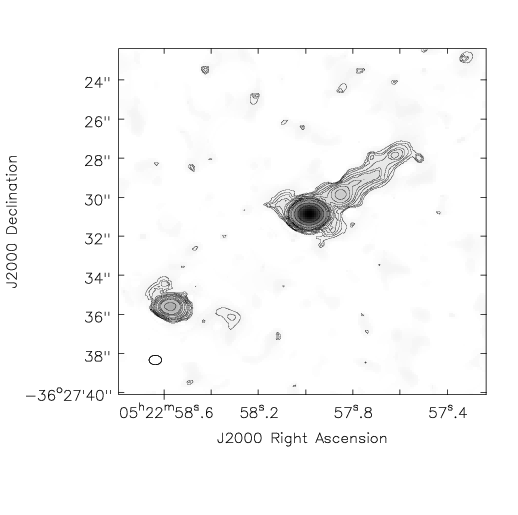}
\caption{Continuum emission towards \pks\ in band 7 ($\nu_{eff} = 336.44$ GHz) after filtering at 4.5-$\sigma$ using a WT. The contours are 
at [3, 4, 5, 7, 9, 11, 15, 20, 30, 40, 50, 100, 200, 1000, 5000]-$\sigma$ for a RMS noise of 91 \mujy.}
\label{fig:B7-wt}
\end{figure*}

\begin{figure*}
\includegraphics[width=12cm]{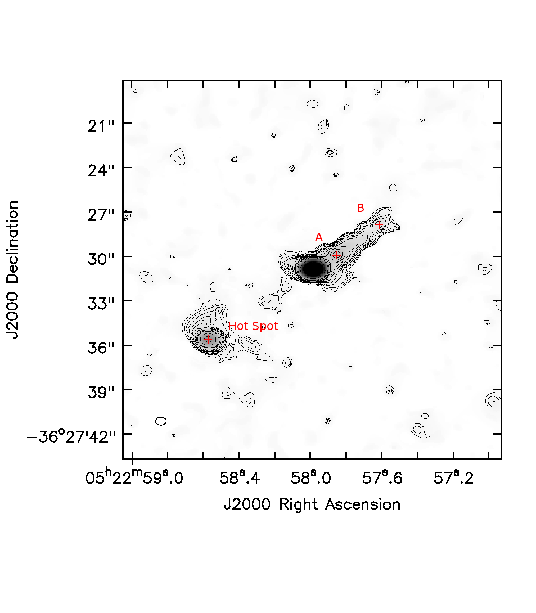}
\caption{Continuum emission towards \pks\ combining the band 3,6 and 7 data ($\nu_{eff} = 221.02$ GHz) after filtering at 4.5-$\sigma$ using a WT. The contours are
at [3, 4, 5, 7, 9, 11, 15, 20, 30, 40, 50, 100, 200, 1000, 5000]-$\sigma$ for a RMS noise of 90 \mujy. The red crosses point on 3 regions indicated in the text.}
\label{fig:B367-wt}
\end{figure*}

\noindent In Figures \ref{fig:B3-wt}, \ref{fig:B6-wt}, \ref{fig:B7-wt}  the WT-filtered emission is shown towards \pks~ in band 3 (100 GHz), band 6 (237 GHz) and band 7 (336 GHz).
Figure \ref{fig:B367-wt} shows the full WT-filtered combined dataset in band 3+6+7. The radio continuum emission shares common features in the three bands: a strong core 
emission, a radio jet oriented in the N-W direction with a $ PA \sim 126^\circ$, a Hot Spot in the symmetric direction of the jet and an elongated jet-like feature close 
to the nucleus and nearly perpendicular to the main jet with a $PA \sim 66^\circ$. That latter feature is more visible in band 6 and 7 than in band 3 mainly because 
of the better  spatial resolution. The main jet has a length of 6.8\arcsec, i.e. 7.7 kpc in projection at the distance of \pks. The jet-like feature has a quite symmetric
feature with a length of about 2.6\arcsec (2.9 kpc) in each direction.
In band 6  the counter-jet is visible towards the S-E with a length of 3.9\arcsec (4.4 kpc). It is probably visible as well in band 3 but the poorer spatial resolution 
is not good enough. Because of the dimming at high frequency the counter-jet is not detected so clearly in band 7. We discuss below the comparison between the jet and 
counter-jet. In band 6 and 7 the Hot Spot is spatially resolved with some extended emission as it was checked by removing a Gaussian component fitted on the Hot Spot and
is indicated by the FWHM of the source \#36 detected by SExtractor (see Table \ref{tab:ps-properties}).
Inside the jet two main spots, dubbed A and B, are visible and were already detected in different 
wavelengths (Tingay \& Edwards 2002, and references therein). 
Together with the HS we give their coordinates
in Table \ref{tab:regions-coord} and their positions in Figure \ref{fig:B367-wt}. We discuss hereafter the diffuse emission outside of the nucleus, jet and HS which 
should trace the interaction of the jet with the local interstellar medium.

\begin{table*}[htbp]
\caption{Coordinates (J2000) of some regions in \pks. See Figure \ref{fig:B367-wt}.}
\centering
\begin{tabular}{|l|l|l|}
\hline
Region  &  RA   & Dec  \\
        & (J2000) & (J2000) \\
\hline
Hot Spot (HS) & 05:22:58.58 &  -36:27:35.6 \\
A             & 05:22:57.85 &  -36:27:29.9 \\
B             & 05:22:57.62 &  -36:27:27.9 \\
\hline 
\end{tabular}

\label{tab:regions-coord}
\end{table*}

\begin{figure*}
\begin{center}
\includegraphics[width=6.5cm]{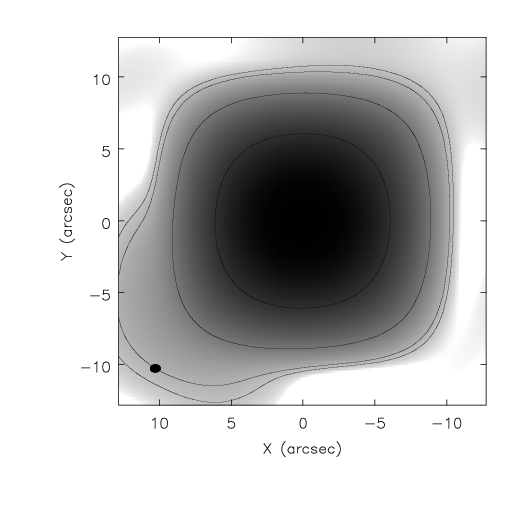}
\includegraphics[width=6.5cm]{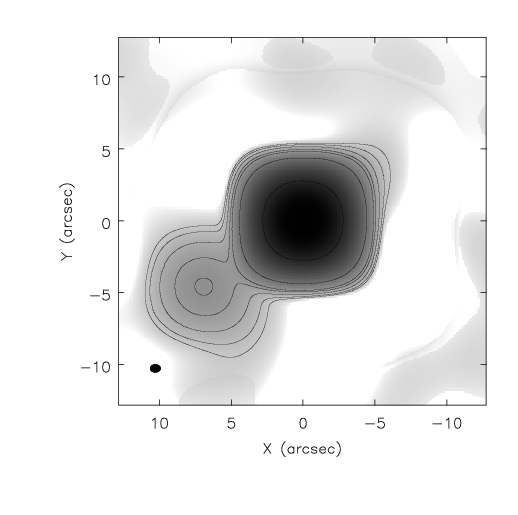}
\includegraphics[width=6.5cm]{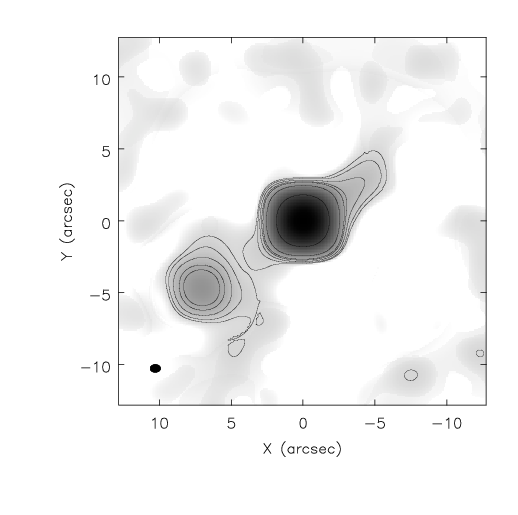}
\includegraphics[width=6.5cm]{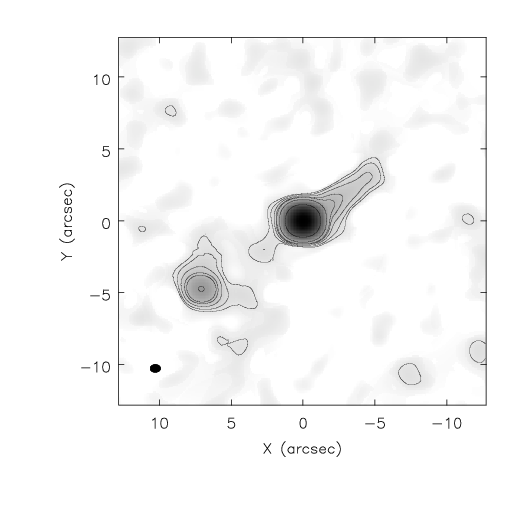}
\includegraphics[width=6.5cm]{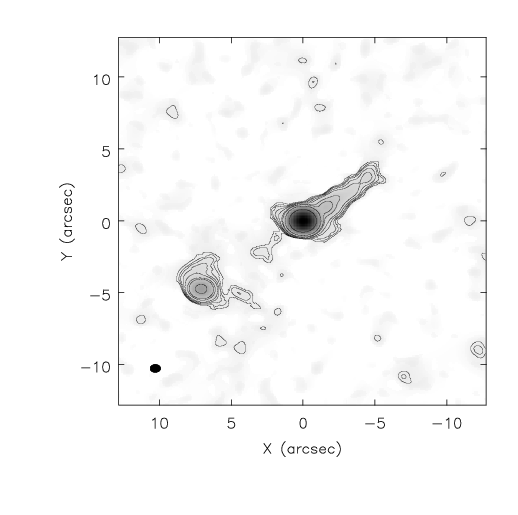}
\includegraphics[width=6.5cm]{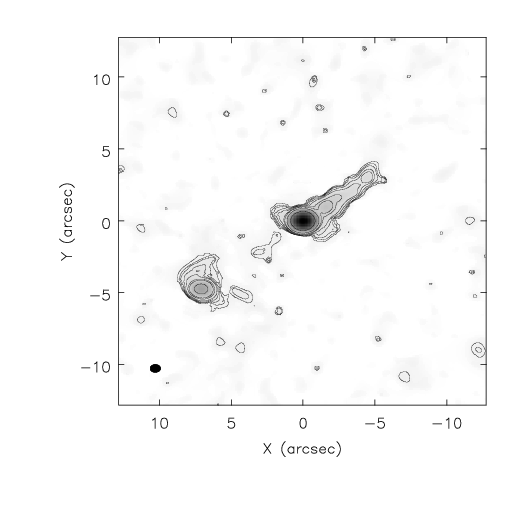}
\end{center}
\caption{Image restoration of the combined band 3+6+7 emission using different range  of the   wavelet filtered planes: planes 6-to-7 (low spatial frequency) 
on the top left figure to planes 1-to-7 (high spatial frequency) on the bottom right figure. Note that the full reconstruction is achieved with the planes 0-to-7. }
\label{fig:B367-wt-panel}
\end{figure*}

\noindent As indicated before, the WT-filtered decomposition allows to analyze the emission present at specific scales. On Figure \ref{fig:B367-wt-panel} the emission in
band 3+6+7 at different scales is restored using the WT-filtered decomposition. The most emphasized features using that technique are the counter-jet 
emission and the  extended emission of the HS. The discrete emission outside the central part is emphasized too in the different reconstructions 
and its relevance was partially addressed before  for the low brightness emission in band 6 and 7 using the wavelet planes 2 to 4 (see Figure \ref{fig:B6B7-2-4})
which select the spatial range 0.4\arcsec-1.4\arcsec (see Leon et al., 2000), i.e. 450-1600 pc in the galaxy \pks.

\noindent Because of the small viewing angle of the radio jet in blazar and in particular in \pks~ the relativistic beaming enhances the approaching blobs of plasma
and dims the receding plasma. The counter-jet is then more visible at low frequencies as it is observed here in \pks~ (see Table \ref{tab:flux-counterjet}). 
Because of the low spatial resolution in band 3 we searched for the counter-jet emission by using 
uniform weights during the imaging step achieving detection in only 3 datasets.
In band 6, 4 datasets reveal the counter-jet emission and because of its low brightness, 
the uncertainty on its flux is quite large. 
To compute the jet/counter-jet emission ratio we use the band 6 estimation with a better spatial resolution, 
i.e the band 3+6+7 dataset which
have an effective frequency in band 6 and a low RMS noise. 
The jet and counter-jet flux was estimated by subtracting to the image a Gaussian source fitting
the nucleus and selecting a polygonal area on the jet and the counter-jet. 
The emission ratio between the jet and the counter-jet was found to be $\Re = 10.2$.

\begin{figure}
\includegraphics[width=9.5cm]{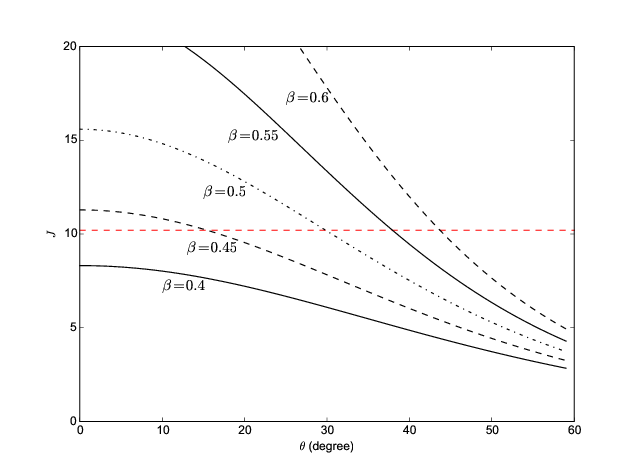}
\caption{Jet/counter-jet ratio $J$ for an exponent $p= 2.5$ in function of the viewing angle $\theta$. The curves for different bulk velocity  $\beta$ are plotted. The 
horizontal line (dashed red line) indicates the flux ratio in \pks.}
\label{fig:jet-counterjet-atio}
\end{figure}

\noindent Assuming an equal intrinsic emission between the jet and the counter-jet, the ratio $J$ between both componet can be estimated as following (e.g. in Ghisellini et al. 1993):

\begin{equation}
J = \left (  \frac{1+ \beta \cos \theta}{1 -  \beta \cos \theta}      \right ) ^ p
\end{equation}
where $\beta$ is the bulk velocity expressed in units of the speed of light, $\theta$ is the viewing angle and $p$ can be expressed in the simplest cases as $p = 2 + \alpha$ 
(a moving isotropic source) or $p = 3 + \alpha$ (a continuous jet) with $\alpha$ the spectral index of the jet. 
On Figure \ref{fig:jet-counterjet-atio} the ratio $J$ is shown for different bulk velocities taking an intermediate exponent $p = 2.5$ which is used for the analysis.
To match the \pks~ ratio it appears that
the viewing angle must be in the range $\theta \subset [16^\circ,38^\circ]$ with a moderate bulk velocity $\beta \subset [0.45, 0.55]$. These values give us a Lorentz
factor $\Gamma \subset [1.1,1.2]$ and a Doppler factor $\delta \subset [1.3,1.7]$.

\noindent These results are similar to the values obtained by Pian et al. (1996) 
which have estimated the Lorentz factor of the plasma responsible for the radio jet to be 
$\Gamma = 1.2$ 
and a viewing angle of $\theta \sim 30^\circ$ giving a Doppler factor of $\delta = 1.6$.
It is also consistent with the space parameter exploration of Tingay et al. (2002) (See Figure 5
in their work).

\begin{table*}[htbp]
\caption{Flux of the counter-jet in \pks.}
\label{tab:flux-counterjet}
\centering
\begin{tabular}{|l|l|l|l|l|}
\hline
band   & $\nu_{eff}$  &  Flux       &  RMS      &  Comments  \\
       &   (GHz)      &   (mJy)     & (mJy) &             \\
\hline
  3    &   106.62      &  15.0      &   1.0    &  J0522-364\_99\_3 uniform weighting   \\
  3    &   104.33      &  20.0	    &   3.9    &  J0522-364\_99\_3 uniform weighting   \\
  3    &   102.40      &  25.5      &   4.7    &  J0522-364\_467\_1 uniform weighting  \\
\hline
  6    &   221.02      &  0.8       &   0.16   &  band 3,6,7 combined and WT filtered \\
  6    &   237.06      &  1.8       &   0.26   &  band 6 combined and WT filtered \\     
  6    &   225.43      &  1.2       &   0.3    &  J0522-364\_210\_1\_2  \\
  6    &   225.43      &  3.1       &   0.7    &  J0522-364\_210\_3      \\
\hline
  7    &    -          &    -       &    -     &      Not detected       \\
\hline
\end{tabular}

\end{table*}

\begin{figure}
\includegraphics[width=9.5cm]{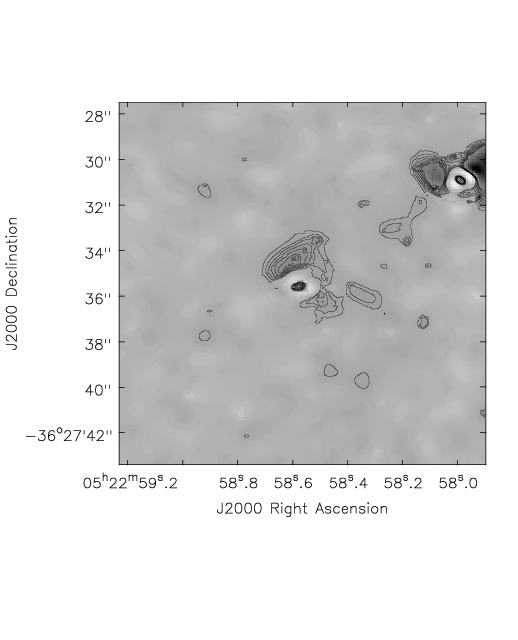}
\caption{Emission of the full data set in band 3, 6 and 7. The contours are at [3,4,5,6,7,8,9,10]-$\sigma$ for a RMS noise of 90 $\mu$Jy. 
A Gaussian was fitted
and removed from the core emission. The counter-jet is visible at the South-East between the HS and the core.}
\label{fig:hs-gaussian-removed}
\end{figure}

\noindent The HS is located at 8.6\arcsec\  from the core position, or 9.9 kpc in projection. Assuming a viewing angle of $30^\circ$ for the jet we estimate the distance from
the center to be 19.8 kpc. The HS exhibits an extended structure, especially in bands 3 and 7 (see Figure \ref{fig:B3-wt} and \ref{fig:B6-wt}). Removing a fitted Gaussian
on the HS, the extended structure is extended mainly towards the North direction (see Figure \ref{fig:hs-gaussian-removed}), with a projected position angle similar to the jet.

\begin{table}[htbp]
\caption{Flux of regions A and B.}
\centering
\begin{tabular}{|l|l|l|l|}
\hline
Id  &  $\nu_{eff}$  &  Flux (A)  & Flux (B) \\
    &    (GHz)      &   (mJy)    &  (mJy)      \\
\hline
B367-combined	     & 221.02 &   5.0   & 2.3     \\
B6-combined	     & 237.06 &   3.3   & 1.6     \\
J0522-364\_268\_2    & 247.08 &   -     & 1.9    \\
J0522-364\_210\_1\_2  & 225.43 &  3.6   & 1.6    \\ 
J0522-364\_210\_3    & 225.44 &   4.4   & -      \\ 
\hline
B7-combined	     & 336.44 &  5.3   &   2.1     \\ 
J0522-364\_108\_2    & 348.58 &  -   &   1.6       \\ 
J0522-364\_397\_1    & 344.99 &  -   &    1.7   \\
J0522-364\_124\_1\_2  & 324.15 & 4.0 &    -     \\
J0522-364\_768\_2\_3  & 345.77 & 5.0 &  1.6     \\    
J0522-364\_83\_1     & 348.82 &  3.3  &    -      \\   
J0522-364\_208\_1    & 348.42 &  3.9  &   -       \\  
J0522-364\_223\_1    & 338.19 &  6.1  &   -       \\   
J0522-364\_223\_2    & 338.19 &  6.7  &   -      \\  
J0522-364\_340\_1\_2   & 365.28 & 7.0   &         \\

\hline
\end{tabular}
\label{tab:flux-A-B}
\end{table}

\section{(Sub-)millimeter variability}
\label{var}
\begin{figure}
\includegraphics[width=10cm]{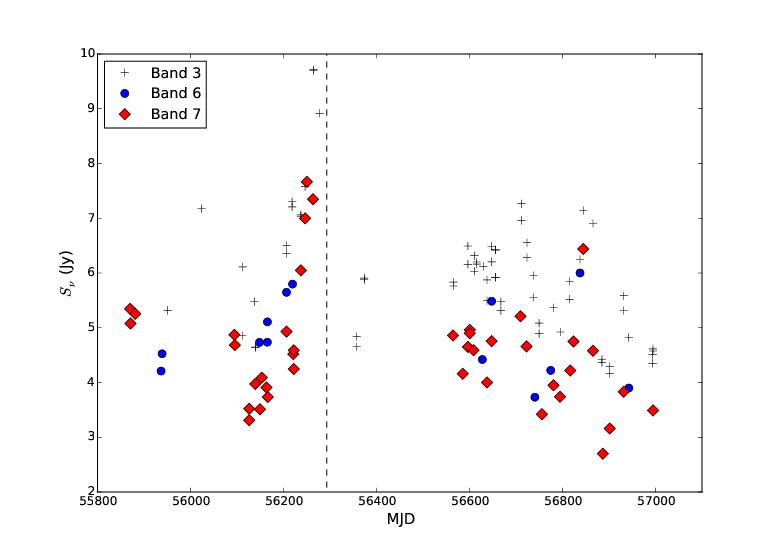}
\caption{Flux evolution  of the core of \pks. For the data from the Source Catalogue we removed the Hot Spot flux using the spectral fit.}
\label{fig:J0522-fluxtime}
\end{figure}

The flux evolution  of \pks~ in three bands of ALMA, 3, 6 and 7, were measured during  nearly 1 year using calibration data of several unrelated scientific projects. The 
time sampling for each band was irregular and did not allow to study completely the very fast variation 
(intra-days) or to measure a precise time variation of the individual blobs in the jet given the low signal to
noise of the individual measurements, e.g. using 
a Fourier decomposition in the time domain. 
Nevertheless, it was possible to disclose important flux variation. 
On Figure \ref{fig:J0522-fluxtime} the evolution of the  
total flux in the three bands is shown. No correction for the frequency was applied but given the
spectral index of the core at the different bands (see Table \ref{tab:J0522-mean}) the maximum correction would be in band 7 of the order of 4\%.

\noindent The variability index VI (Ciaramella et al., 2004) was introduced to parametrize the flux variation of a source:
\begin{equation}
 VI = \frac{S_{max} - S_{min}}{S_{max}+S_{min}}
\end{equation}

\noindent For each band, the VI of \pks~  is then estimated on the sampling period of 1 year to be respectively  $VI_{B3} = 0.33 $, $VI_{B6} =  0.06$ and $VI_{B7} = 0.40 $ 
(see Table \ref{tab:J0522-mean}). These values of the VI must be considered as lower limit during the year 2012 since the sampling is sparser than the variability 
in \pks. 
The sparse and irregular sampling makes difficult to draw clear conclusions on the 
variability between the bands. Nevertheless, it seems that the emission in band
3, the lowest frequencies,  is more variable than the other two bands.

\section{Spectral index}
\label{spindex}
\begin{figure}
\includegraphics[scale=0.5]{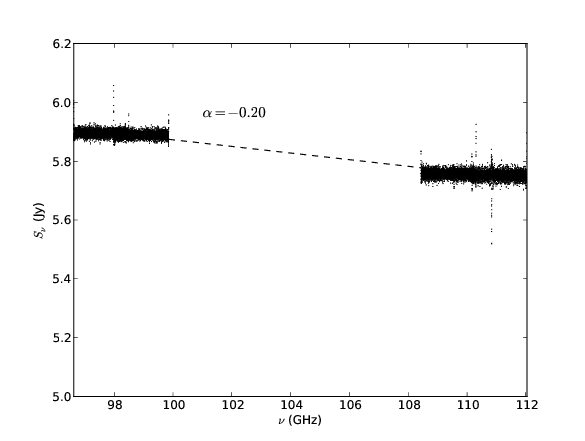}
\caption{Spectral index using the 4  simultaneous basebands observed with ALMA. Here is the example of the set J0522-364\_99\_2. The spectral index
fit is indicated by the dashed line.}
\label{fig:99_2-intrabandSI}
\end{figure}

\begin{figure}
\includegraphics[width=10cm]{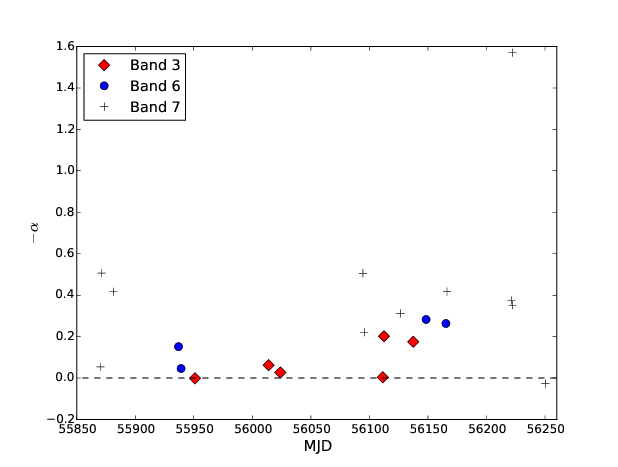}
\caption{Evolution of the spectral index $\alpha$ in the core of \pks\ vs. MJD. The data from the calibrator monitoring were added to the
project data. In that case we remove the Hot Spot flux estimated from the  SED fit.}
\label{fig:J0522-sitime}
\end{figure}

On Table \ref{tab:obs-SI} 
the spectral index is given per dataset. The spectral index was computed measuring  the core flux at the 4  basebands of a given dataset (see Figure \ref{fig:99_2-intrabandSI}).
The mean value of the spectral index was computed for each band (see Table \ref{tab:J0522-mean}) and shows a clear correlation with the frequency with a steepening 
towards the high frequency, $-0.43$ in band 7 versus $ -0.08$ in band 3. The time evolution of the spectral 
index for each band (see Figure \ref{fig:J0522-sitime}) shows a very quick
change for the spectral index in band 7, which in less than 5 days it can change by more than 0.5 dex. 
That behavior was already shown at lower frequencies in other blazars (e.g. Venturi et al., 2001).

\begin{figure}
\includegraphics[width=10cm]{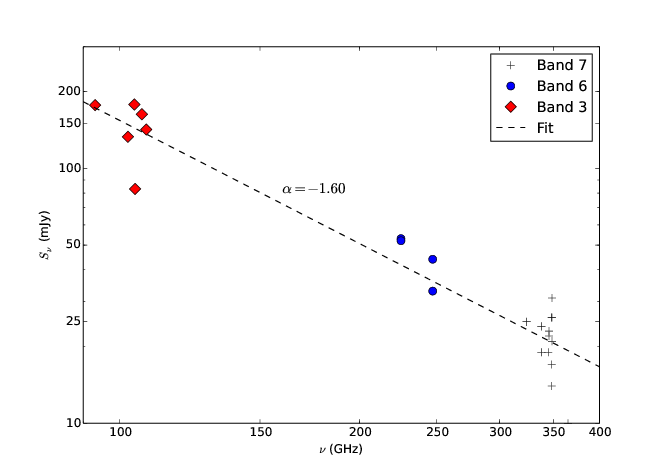}
\caption{SED of the Hot Spot in band 3, 6 and 7 and the fit of the spectral index $\alpha = -1.60$.}
\label{fig:hotspot-sed}
\end{figure}

\begin{figure}
\includegraphics[width=10cm]{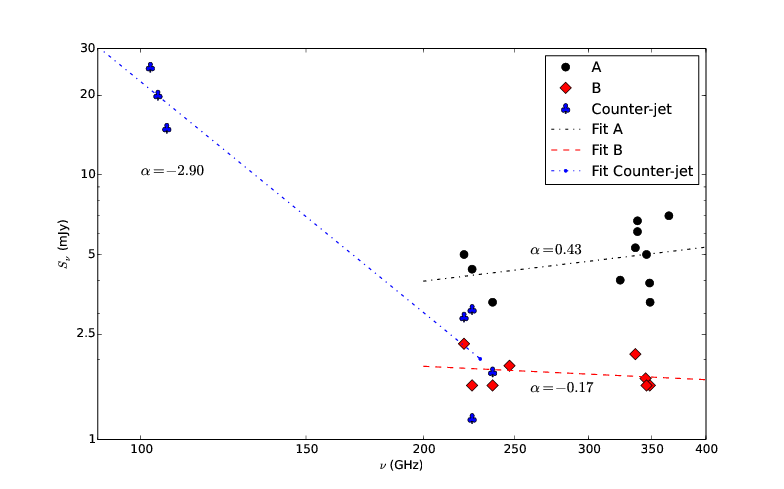}
\caption{SED of the measured continuum data in the regions A, B and the counter-jet. A spectral index is fitted on the three regions and indicated in the plot.}
\label{fig:SED-regions}
\end{figure}

\noindent The dynamic range achieved for the imaging of the different  dataset allows to measure 
the flux of the other components. Since the regions A and B are embedded in the jet emission we take a fixed circular aperture of 1\arcsec around the 
peak emission to estimate the flux. The SED of these components
is shown on Figure \ref{fig:hotspot-sed} (HS) and \ref{fig:SED-regions} (counter-jet, A and B).  The very low value, $-2.9$, of the spectral 
index of the counter-jet is 
not only reflecting the intrinsic flux distribution but as well the relativistic beaming of 
the jet/counter-jet as discussed before.  The HS discloses a  steep distribution
with a spectral index of $-1.6$ similar to Hot Spots at high frequency in other radio galaxies (Carilli et al. 1999, Okayasu et al., 1992).

\noindent The regions A and B in the radio jet have a  different behavior with a flat or even inverted spectrum. We note that because of 
the spatial resolution necessary
to discriminate the regions A and B the spectral index was measured from the  band 6 and 7. Figure \ref{fig:SED-regions} makes clear the large dispersion of the 
measurements. The time variation cannot be discarded even if it is not expected. 
A most likely reason for these variations is the difficulty to extract the flux 
from a  region with a single dataset because of the low dynamic range in the radio jet and the 
ill-defined location. 
Despite these issues we can  confidently affirm
that regions A and B have a relatively flat spectrum in band 6 and 7.

\begin{table*}[htbp]
\caption{Spectral index of the core of \pks\ measured on the basebands.}
\centering
\begin{tabular}{|l|l|l|}
\hline
Id  &  $\nu_{eff}$  &    $\alpha$  \\  
    &    (GHz)      &             \\
\hline
J0522-364\_99\_1      & 104.55 &     -0.03 \\
J0522-364\_99\_2      & 104.33 &     -0.20  \\
J0522-364\_99\_3      & 106.62 &      0.0   \\
J0522-364\_131\_1     & 107.99 &      0.00  \\
J0522-364\_467\_1     & 102.40 &      -0.06   \\
J0522-364\_108\_3     & 93.16  &      -0.18    \\
\hline
J0522-364\_268\_1     & 247.08 &     -0.15       \\
J0522-364\_268\_2     & 247.08 &      -0.05     \\
J0522-364\_210\_1\_2  & 225.43 &       -0.28     \\ 
J0522-364\_210\_3     & 225.44 &       -0.26        \\ 
J0522-364\_108\_1     & 348.58 &       -0.05     \\
J0522-364\_108\_2     & 348.58 &     -0.51                \\
J0522-364\_397\_1     & 344.99 &      -0.42       \\
J0522-364\_124\_1\_2  & 324.15 &      -0.31      \\
J0522-364\_83\_1      & 348.82 &      -0.51    \\   
J0522-364\_83\_2      & 348.82 &      -0.22      \\  
J0522-364\_83\_3      & 348.82 &       -0.42       \\  
J0522-364\_208\_1     & 348.42 &       -1.57    \\ 
J0522-364\_223\_1     & 338.19 &       -0.37      \\   
J0522-364\_223\_2     & 338.19 &       -0.35      \\  
J0522-364\_340\_1\_2  & 365.28 &         0.03    \\ 
\hline
\end{tabular}

\label{tab:obs-SI}
\end{table*}

\begin{table}[htbp]
\caption{Mean values of  core spectral index in \pks\ and the VI per band without correcting for the spectral dependence.}
\centering
\begin{tabular}{|l|l|l|l|}
\hline
band   & \#  &   $\alpha$  &  VI \\
\hline
3      & 6   & -0.08      &    0.40    \\
6      & 4   & -0.19      &    0.23     \\
7      & 13  &  -0.43      &    0.48    \\
\hline
\end{tabular}

\label{tab:J0522-mean}
\end{table}

\section{Discussion}
\label{discussion}

\subsection{Beamed radio jet}
The detection of the jet and  counter-jet emission allows to use 
their ratio, $\mathfrak{R}$, as an alternative  beaming indicator assuming no isotropic
component in the jet emission (Ghisellini et al. 1993). 
This beaming indicator is independent of other 
methods such as the proper motion of blobs and the
core to extended radio flux usually used with  VLBI observations.
Also, the resolution required to detect kinematic motions within the jet by ALMA is 
difficult to achieve frequently enough to monitor the movement of blobs within the jet.
But taking advantadge of the ALMA sensitivity, the detection of the counter-jet by 
the stacking approach followed here is a promising avenue at these frequencies.
Our derived value for $\mathfrak{R}$ is similar 
to those values derived  from previous VLBI observations. 
Tingay et al. (1996) found lower limits for $\mathfrak{R}$ in the range of our findings (5.0
and 9.3 at 4.8 GHz, and greater than 2.9 at 8.4 GHz).  
Further observations by Tingay \& Edwards (2002) 
obtained $\mathfrak{R}>20$ from the VLBI observations at 5  and 8.4 GHz increasing the
previously obtained lower limit given the higher sensitivity of the new VLBI data.
Although our result differs in a factor of 2, it is still within the expected range 
when compared to other highly beamed blazars (Ghisellini et al. 1993).
As previously described, from $\mathfrak{R}>20$ we derived bounded ranges for $\delta$, 
$\Gamma$, $\beta$, and $\theta$ (see Section \ref{ana}) which mean values are in good
agreement with previous studies, where $ \langle\delta\rangle = 1.5$, $ \langle\Gamma\rangle = 1.15$, 
$\langle\theta\rangle = 32.5^{\circ}$, and $\langle\beta\rangle = 0.5$.
In a highly beamed blazar, the ratio $\mathfrak{R}$ get enhanced due to the beaming effect in the 
jet. Thus, a small value of  $\mathfrak{R}$ suggests that we are seeing the true bulk motions
in the jet and not emission due to beaming effects.
Tingay \& Edwards (2002) observations constrain the parameter space to a well defined region 
which centered the jet angle close to $30^{\circ}$ and $\beta \sim 0.5$.
These exploration of the $\beta - \theta$ plane also considered the
calculations done by Ghisellini et al. (1993) for the Doppler parameter ($\delta > 1$) and 
Dondi \& Ghisellini (1995) ($\delta > 1.3$),
which we can now further constrain with our ALMA data.
It is no coincidence that our derived parameters lie just 
inside the box allowed by Tingay \& Edwards (2002) in the $\beta - \theta$ plane and is also 
consistent with the values used for $\delta$.
Therefore, we found little support for highly beamed emission in \pks~ and thus, 
though not detected here, it is unlikely that there are super-luminal motions along the jet.

\subsection{Double jet, dust or free-free emission}

One of the striking results present in the millimeter continuum emisson towards \pks~ is an elongated 
structure nearly perpendicular to the 
well-defined radio jet (see Fig. \ref{fig:B367-wt}). In the combined dataset with an effective frequency of 
221 GHz that structure has a size of 4.8\arcsec, i.e. 
5.5 kpc at the distance of \pks, and a position angle of $72^\circ$. The isophote fitting of the R-band 
image (Falomo, 1994) indicates that this feature 
is aligned with the major axis of the host galaxy (PA$\approx 75^\circ$). The thermal (dust) emission or free-free emission from central star
forming region or
synchrotron emission are then possible explanations for that
feature. Previous observations have shown the presence of kpc-scale disk of molecular gas and dust in 
the center of radio galaxies (Martel et al. 1999, 
Garc\'{i}a-Burillo et al. 2007, Oca\~na Flaquer et al. 2010). It has been observed as well X-shaped radio sources with normal pairs of 
active lobes and lower surface brightness wings of emission (Leahy \& Parma 1992, Dennett-Thorpe et al. 2002, Cheung 2007). 
The 15 GHz VLA map shown in Falomo et al. (2009) has a similar angular resolution ($\sim 0.5\arcsec$) to the ALMA data and a RMS noise of 
200 \mujy\ which is sufficient to detect the perpendicular feature if it is a synchrotron emission. No hint of an elongated feature along the major axis is 
visible on the 15 GHz map where all the other features are clearly detected (HS, jet).
Boisson et al. (1989) found a nebular line emission with a low surface brightness extending up to 21 kpc on the opposite side of the radio jet. 
It does not match spatially with the secondary emission found in band 6 and 7 with ALMA. The spectral index  of the emission in the elongated feature between the band 3 and 7 
would allow to discriminate between synchrotron and thermal  emission since in most of the case the SED slope are negative and positive respectively. The spatial 
resolution of the current band 3 data does not allow to resolve that feature. The lack of evidence of that feature at 15 GHz favors the dust emission but it will need to 
be confirmed by observations in band 3 at a higher spatial resolution. In that case the presence of dust in the center of \pks~ should be associated with molecular gas that 
we should be able to detect in CO(1-0) emission or other tracers (Oca\~na Flaquer et al. 2010).

\section{Conclusions}

\label{conclusions}
 The blazar \pks~ was studied using the ALMA public archive from the projects using that source as a calibrator over more than 11 months. The data in band 3, 6 and 7 were
 analyzed separately to follow the time variability and combined for each band in a 
superset at an effective frequency of 221 GHz. To enhance the imaging we applied
 a wavelet filtering using the "A trous" algorithm to reach a final dynamic range of 47000 in the superset of data. From that analysis we can draw several conclusions
 about the (sub-)millimeter emission of \pks :
 
 \begin{itemize}
  \item The detection of the submillimeter point sources in the field of \pks~ are consistent with sub-millimeter galaxy count obtained with similar data using ALMA observations. They show 
  the strong potential of the numerous calibrator fields used in ALMA observations to build a large sample of high-redshift sources over the whole sky spanned by the  
  ALMA scientific projects. A careful analysis of the sources, beyond the scope of this study, would be needed to discard spurious detections.
  \item The (sub-)millimeter emission of the extended structures in \pks~ is detected in band 3, 6 and 7. The Hot Spot was clearly detected at  8\arcsec\ from the core.
  The strongly beamed radio jet was detected together with its internal regions (A and B). We detected the presence of a weak counter-jet, symmetric to the radio jet. 
  A roughly perpendicular disky structure is detected in band 3 and 6. That structure may be  the relic of a previous jet but a thermal origin cannot be discarded without
  more observations at higher spatial resolution and frequency.
  \item The analysis of the time variability shows large variations during the nearly 1 year of measuremens. The irregular sampling makes difficult to perform a frequency analysis of
  the variation. The spectral index variation at the three bands shows large variation reaching 0.4-0.6 dex difference.
  \item The spectral index analysis indicate a steepening of the SED in the core towards the highest frequencies (band 7) whereas the regions A and B have a quite flat 
  spectrum. The counter-jet displays a very steep SED because of the relativistic beaming in the blazar. Analyzing the data of the jet and counter-jet we were able to
  put constraint on the Lorentz factor and the viewing angle confirming previous studies at lower frequencies. According to that analysis the Hot Spot would be at a distance
  of nearly 20 kpc from the core.
  \item The Hot Spot shows an extended structure apart from its strong central emission. Despite the lower spatial resolution in band 3 it seems to present more extended 
  structures towards the low frequencies.
  \item Our derived parameters for $\langle\delta\rangle = 1.5$, $\langle\Gamma\rangle = 1.15$,
$\langle\theta\rangle = 32.5^{\circ}$, and $\langle\beta\rangle = 0.5$, are within the range of the constraints imposed by 
VLBI observation in the $\beta - \theta$ plane indicating that the emission from \pks~
is definitely not highly beamed and thus, support minimal super-luminal motions along the jet. 
  
 \end{itemize}

\begin{acknowledgements}
      We are grateful to the anonymous  referee  for the very useful comments. 
      M. G. thanks ESO for his studentship. This paper makes use of the following ALMA data: ADS/JAO.ALMA\#Id (Id are the project UID indicated in Table \ref{tab:obs-journal}).
      ALMA is a partnership of ESO (representing its member states), 
      NSF (USA), and NINS (Japan), together with NRC (Canada) and NSC and ASIAA (Taiwan), in cooperation with the Republic of Chile. The Joint ALMA Observatory 
      is operated by ESO, AUI/NRAO, and NAOJ.
\end{acknowledgements}

\bibliographystyle{aa}

\end{document}